# Multiresolution Analysis and Learning for Computational Seismic Interpretation

Motaz Alfarraj, Yazeed Alaudah, Zhiling Long, and Ghassan AlRegib

Georgia Institute of Technology, Atlanta, GA

## ABSTRACT

We explore the use of multiresolution analysis techniques as texture attributes for seismic image characterization, especially in representing subsurface structures in large migrated seismic data. Namely, we explore the Gaussian pyramid, the discrete wavelet transform, Gabor filters, and the curvelet transform. These techniques are examined in a seismic structure labeling case study on the Netherlands offshore F3 block. In seismic structure labeling, a seismic volume is automatically segmented and classified according to the underlying subsurface structure using texture attributes. Our results show that multiresolution attributes improved the labeling performance compared to using seismic amplitude alone. Moreover, directional multiresolution attributes, such as the curvelet transform, are more effective than the non-directional attributes in distinguishing different subsurface structures in large seismic datasets, and can greatly help the interpretation process.

# INTRODUCTION

Texture attributes are mathematical quantities computed to capture the perceived spatial patterns of an image. Such attributes have been employed in some seismic interpretation applications primarily because of the textural content of migrated seismic images. For instance, texture-based approaches have been proposed for seismic image segmentation (e.g. Pitas and Kotropoulos, 1992; Röster and Spann, 1998), salt body detection (e.g. Hegazy and AlRegib, 2014; Shafiq and AlRegib, 2016), seismic structure labeling (Alaudah and AlRegib, 2016), and seismic image similarity and retrieval (e.g. Al-Marzouqi and AlRegib,2014; Long et al., 2015; Alaudah and AlRegib, 2015; Mattos et al., 2017). The various successful applications of texture attributes are an indication of their potential in the field of seismic interpretation. Using such attributes, it is possible to automatically characterize subsurface imagery and highlight features of interest to interpreters.

Texture attributes can be categorized into two categories: spatial texture attributes, and frequency-based texture attributes. The texture attributes at the core of this study are based on multiresolution analysis techniques, which are also known as frequency-based attributes. Such attributes utilize the frequency content of an image as opposed to spatial attributes which exploit correlations and statistics in the spatial domain (e.g. Haralick et al., 1973; Ojala et al., 2002; Berthelot et al., 2013; Zhai et al., 2013).

In this study, we explore the Gaussian pyramid (Adelson et al., 1984), which is a multiscale image representation technique that led to the development of multiresolution analysis techniques. Moreover, we explore the classical discrete wavelet transform (DWT), which is one of the most popular techniques for spectral content-based texture analysis. In addition, we examine a multiresolution texture attribute derived from Gabor filters which are linear filters designed to extract edge information at different frequencies and orientations (Randen and Sønneland, 2005). The fourth attribute we examine is the curvelet transform (Candes et al., 2006) which is considered an extension of the wavelet transform to overcome some of its shortcoming such as its limited directionality.

The goal of this study is to investigate the various multiresolution texture analysis techniques for seismic interpretation, and to show the effectiveness of some of these techniques in characterizing subsurface structures. For this purpose, the attributes are examined through a structure labeling case study on the Netherlands offshore F3 block. In structure labeling, a seismic volume is automatically segmented into different regions according to their respective dominant structure, and each region is assigned with its corresponding label or class. The segmentation and label assignment (or classification) are performed on the basis of the aforementioned attributes.

## MULTIRESOLUTION TEXTURE ATTRIBUTES

In this section, we describe four multiresolution decomposition techniques that are commonly used in image processing. Namely, the Gaussian pyramid, the discrete wavelet transform, Gabor filters, and the curvelet transform.

The Gaussian pyramid is a technique that is used to decompose an image into different scales each of which comprises features of similar size. Pyramid decomposition, in general, laid the groundwork for multiresolution analysis techniques. The discrete wavelet transform improved on pyramid techniques by decomposing each scale into horizontal, vertical and diagonal components. Gabor filters introduce directional decomposition by which the different scales are decomposed into more orientations than those obtained from the discrete wavelet transform. It is worth noting that Gabor filters do not form a transform, i.e. the image response to the filters does not fully represent the original image. The curvelet transform is a natural extension of the wavelet transforms to overcome its limited directionality. It decomposes an image into different frequency bands at different scales and orientations. Throughout this section, we will use the seismic image shown in Figure 1 to illustrate image decomposition using the aforementioned multiresolution analysis techniques.

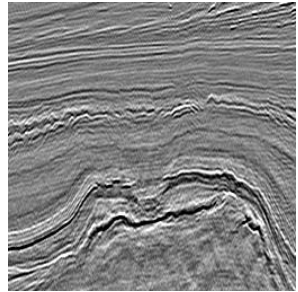

**Figure 1**. Part of a salt dome from inline 299 of the Netherlands offshore F3 block.

**The Gaussian Pyramid**

The Gaussian pyramid is a classical multiscale analysis technique, which is the predecessor of multiresolution analysis techniques. It has been used for various applications in image and video processing such as image coding, image and video compression, and salient object detection (e.g. Burt and Adelson, 1983; Adelson et al., 1984; Itti et al., 1998). In this work, the Gaussian pyramid serves as an efficient multiscale analysis tool to exploit features of different sizes in a seismic image. For a 2D seismic image, $\mathbf{I}_0$, the $k$-scale Gaussian pyramid is constructed as follows. First, $\mathbf{I}_0$ is set as scale 0 of the pyramid which represents the full resolution scale. Then, scale 1 of the pyramid, $\mathbf{I}_1$, is computed by smoothing $\mathbf{I}_0$ with a Gaussian filter, followed by downsampling it by a factor of 2. The remaining scales are generated in a similar fashion. An illustration of a 4-scale

Gaussian pyramid of a seismic section is shown in Figure 2. Note that the dimensions of the image are reduced by a factor of 2 each time a one-level decomposition is performed (i.e., proceeding upward along the pyramid by one level). The Gaussian blurring filter serves as a low-pass filter and is followed by a downsampling step to avoid redundancy.

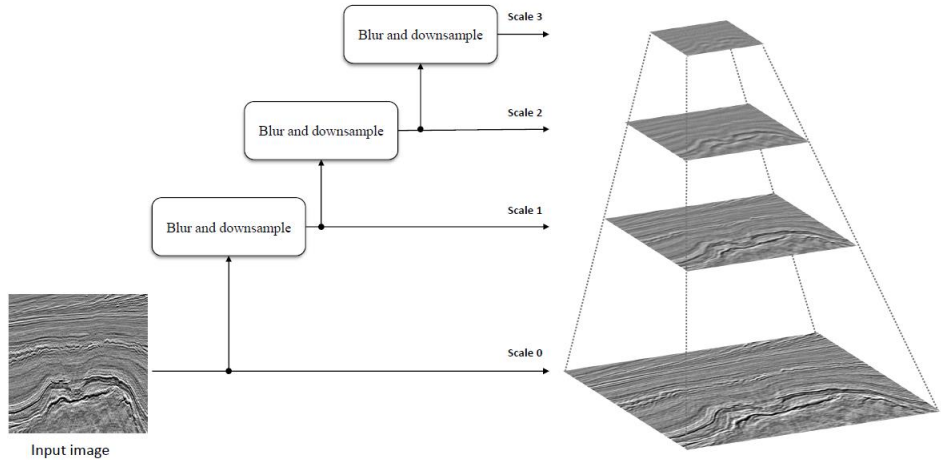

**Figure 2.** 4-scale Gaussian pyramid workflow.

**Discrete Wavelet Transform**

The discrete wavelet transform (DWT) is an orthonormal transform that represents an image using a dyadic dilation and translation of a certain basis function (or a mother wavelet). Different wavelet bases have been proposed and studied extensively such as Haar, Daubechies, symlet, Mexican hat, and coiflet wavelets, among many others (Daubechies, 1992). A classical choice of such basis function is $\mathbf{h}_L = \frac{1}{\sqrt{2}}[1, 1]^T$ and $\mathbf{h}_H = \frac{1}{\sqrt{2}}[-1, 1]^T$, which is known as the Haar wavelet. The first-level discrete wavelet coefficients of an image are obtained by filtering along the horizontal direction with low pass $\mathbf{h}_L$ and high pass $\mathbf{h}_H$ filters to obtain $\mathbf{I}_L^{(1)}$ and $\mathbf{I}_H^{(1)}$ respectively. Then, $\mathbf{I}_L^{(1)}$ and $\mathbf{I}_H^{(1)}$ are filtered along the vertical direction with the same filters and decimated by a factor of

2 to obtain three detail images $\mathbf{I}_{HH}^{(1)}$, $\mathbf{I}_{HL}^{(1)}$ and $\mathbf{I}_{HL}^{(1)}$, and one approximation image $\mathbf{I}_{LL}^{(1)}$ as shown in Figure 3(a). For more levels, the same process is repeated on the approximation image $\mathbf{I}_{LL}^{(1)}$. An example of a 2-level DWT of a seismic image is shown in Figure 3(b). The total number of bands depends on the number of levels. For instance, a 4-level DWT will result in 13 subbands (3 detail bands × 4 levels + 1 approximation band).

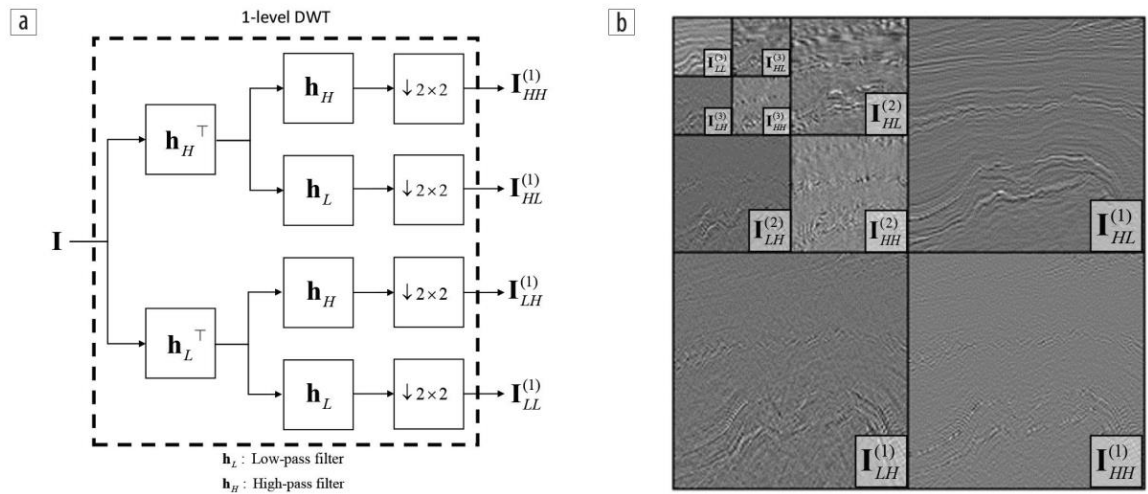

**Figure 3.** (a) 1-level 2D discrete wavelet transform workflow. The same workflow is used to generate the next-level decomposition from the approximation image ($\mathbf{I}_{LL}^{(1)}$). (b) A 3-level discrete wavelet transform of the image in Figure 1.

**Gabor Filters**

Gabor filters are linear filters that are the product of a 2D plane wave with a Gaussian filter. Gabor filters are frequently used as models of the simple cell receptive fields in the human visual system (Daugman, 1988). They have been utilized to characterize natural and texture images especially for applications such as edge detection (Mehrotra et al., 1992) and image segmentation (Jain and Farrokhnia, 1991). Furthermore, Gabor filters have been used in seismic image processing to extract useful characteristic features and define some seismic attributes (Randen and Sønneland,

2005). Figure 4 shows the response of the image in Figure 1 to different Gabor filters at 3 scales and 4 orientations.

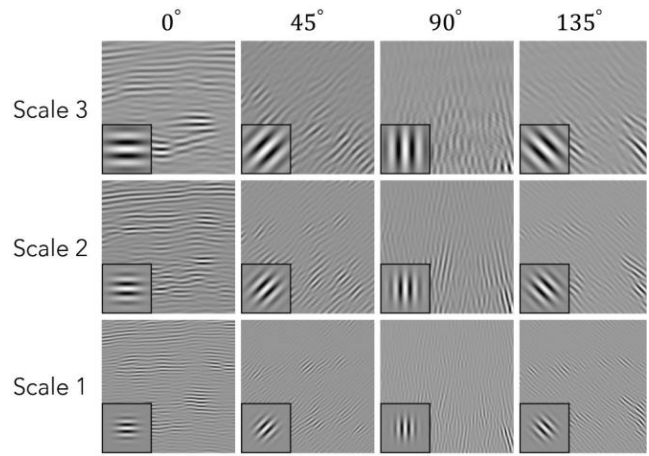

**Figure 4.** The image in Figure 1 filtered with Gabor filters at 3 scales and 4 orientations. The filters are shown at the bottom left corners.

**Curvelet Transform**

The curvelet transform is a multiscale directional decomposition. Despite their popularity, wavelets fail to compactly represent images with highly directional elements such as curves and edges. To the contrary, curvelet frames have been shown to represent images with geometrically regular edges (such as seismic images) more compactly than other traditional multiscale representations (Candes et al., 2006). For an image with $n$ pixels, the fast discrete curvelet transform (FDCT) allows the computation of curvelet coefficients in $O(n\log n)$ operations making the curvelet transform not only fast to compute but also scalable to very large images.

For the purposes of this study, we present a simplified overview of the FDCT. For a detailed description, refer to (Candes et al., 2006). Given an image, the FDCT divides the Fourier support of the image into $J$ scales and $K(j)$ orientations as is shown in Figure 5(a) such that $J = \lceil \log_2 \min(N_1, N_2) - 3 \rceil$, where $\lceil * \rceil$ is the ceiling function. The number of orientations at scale

$j$ is given by $K(j) = 16 \times 2^{\lceil (j-1)/2 \rceil}$. The curvelet coefficients are then obtained by taking the Inverse Fast Fourier Transform (IFFT) of each subband after multiplying it by a smoothing function and wrapping it around the origin. The 3-scale curvelet coefficients of the image in Figure 1 are shown in Figure 5(b). Note that we use generic labels of the axes in Figure 5(a), namely, horizontal and vertical frequency instead of wavenumber and frequency in order to generalize the decomposition for inline/crossline as well as time slices.

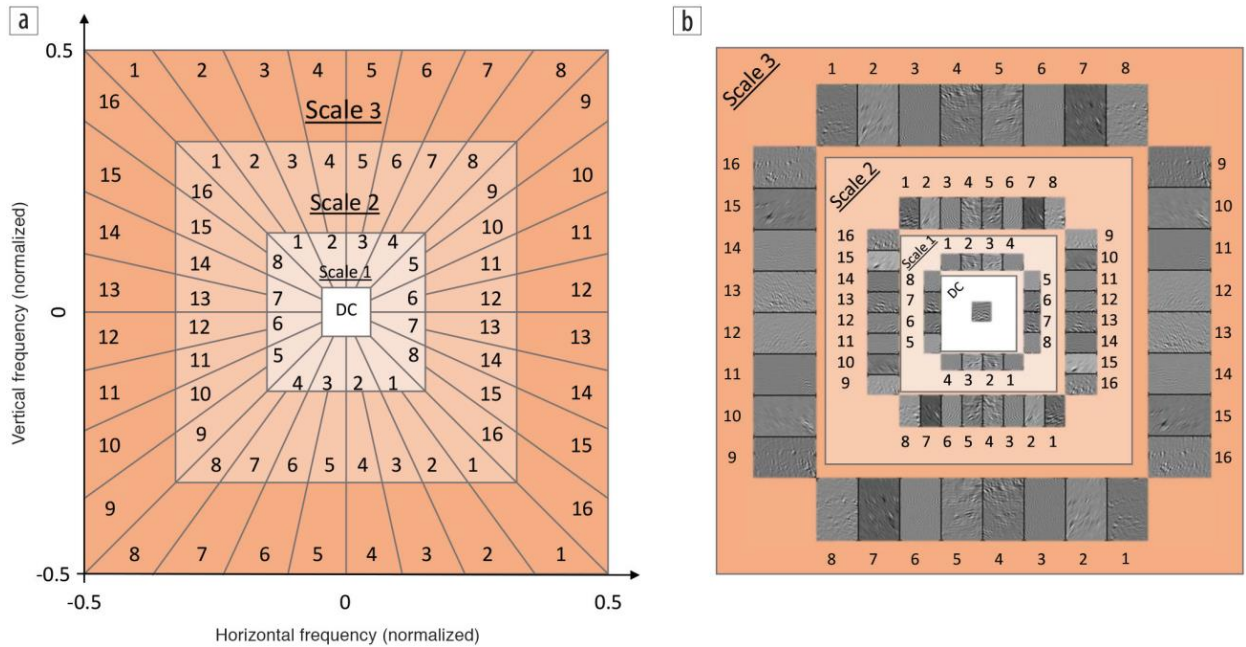

**Figure 5.** (a) Generic frequency domain partitioning of a 3-scale curvelet transform. (b) The real part of the corresponding curvelet coefficients of the image in Figure 1. Orientations and scales are represented by numbers and colors, respectively. Note that the spectrum of a 2D image is conjugate-symmetric which can be observed from the repeated orientation numbering for each scale.

## CASE STUDY ON THE NETHERLANDS OFFSHORE F3 BLOCK

We evaluate the capabilities of the aforementioned multiresolution attributes in characterizing seismic images using structure labeling of the Netherlands offshore F3 block. We use a subset of the publicly available LANDMASS-1 dataset (CeGP, 2015) to form a training dataset.

LANDMASS-1 consists of more than 17,000 images of size 99×99 pixels extracted from the Netherlands offshore F3 block dataset (Opendtect, 1987). These images contain various subsurface structures such as `horizons`, `chaotic`, `faults`, and `salt domes`. From within the LANDMASS-1 dataset, we retrieve images that are most similar to each of the example images shown in Figure 6 according to the similarity measure proposed in (Alfarraj et al., 2016). These example images are hand-selected to exemplify four classes of subsurface structures. Namely, these classes are `chaotic`, `faults`, `salt domes`, and `other`. The `other` class contains all the images that have structures that are not in the first three classes, such as clear horizons and sigmoidal structures. The `other` class serves the purpose of showing negative examples of structures that do not belong to the first three classes. Overall, we retrieve 1000 images for the `other` class, 1500 for `salt domes`, and 500 each for `chaotic`, and `faults`.

Furthermore, to validate our labeling results, we use four manually labeled inline sections from the Netherlands offshore F3 block. Namely, we use inlines number 160, 310, 330, and 380. These sections are shown in Figure 8.

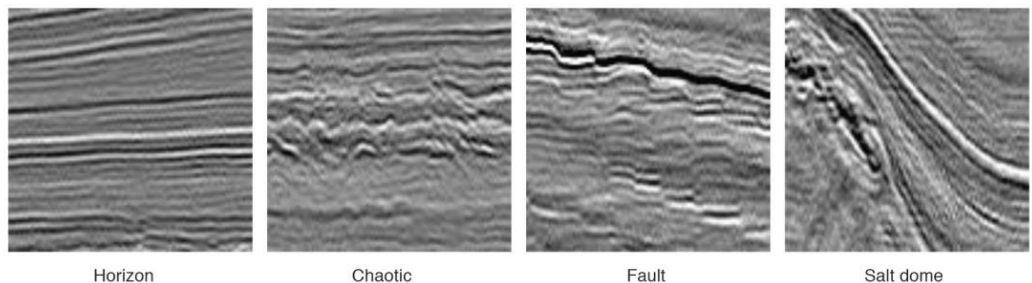

**Figure 6.** Example images of the 4 classes in LANDMASS-1 dataset.

**Structure Labeling Procedure**

The procedure of this case study closely follows the weakly-supervised approach of (Alaudah and AlRegib, 2016). The process is divided into three steps: oversegmentation, feature extraction, and classification. The workflow is depicted in Figure 7.

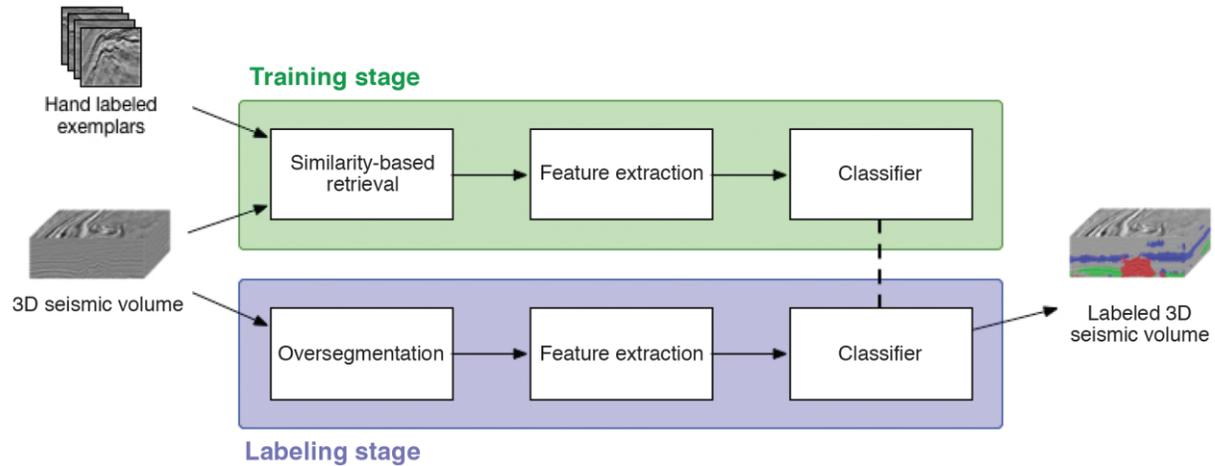

**Figure 7.** Seismic structure labeling workflow.

*Oversegmentation*

Given a large seismic section, we oversegment it into smaller regions using simple linear iterative clustering (SLIC) superpixels (Achanta et al., 2012) to enforce local spatial correlations in seismic data. However, since seismic images are grayscale, we cluster the pixels in the $[l, g_x, g_y, x, y]$ space instead of the original $[l, a, b, x, y]$ space when we apply the SLIC algorithm. Here, $l$ is the seismic amplitude, $g_x$ and $g_y$ refer to the gradient of the seismic section along the $x$ and $y$ directions respectively, and $x$ and $y$ are simply the $x$ and $y$ coordinates of each pixel.

*Feature Extraction*

Given a specific superpixel from a seismic section, we extract an image of size 99×99 centered around the centroid of the superpixel. We then compute the element-wise product of this image with a two-dimensional Gaussian kernel of the same size which is centered at the middle of the image with its variance selected such that pixels in the corners of the image have weights of less than 1%. Thus, more weight is given to the structures at the center of the image.

Each image is then decomposed into subbands of different scales and/or orientations using the attributes detailed in the previous section. A feature vector of the image is formed using a given attribute in two steps. First, we compute the effective singular values of each subband as detailed in (Roy and Vetterli, 2007). Second, all effective singular values of all subbands are concatenated in one feature vector. The length of the vector will depend on the number of subbands and the dimensions of each subband which vary from one attribute to another. In this case study, used a 3-scale decomposition for all multiscale attributes.

*Classification*

We train four one-vs-all support vector machines (SVM) (Vapnik, 1999) with linear kernels using the features extracted from the training images after weighting them with a Gaussian kernel. Later at the labeling stage, images centered at each superpixel are extracted. Then the SVMs are used to predict the label of a superpixel based on its corresponding center image. This is performed for all superpixels in a seismic section until the entire section is labeled.

**Results and Discussion**

We labeled inlines 160, 310, 330, and 380 using the multiresolution texture attributes discussed before, in addition to the seismic amplitude where the features are extracted directly from the image. The quantitative results were computed for all 4 sections, and the average scores are reported in Table 1. The labeled sections are shown in Figure 8.

Table 1. Quantitative evaluation of the labeling results averaged over all 4 inlines.

| Attribute | PA | UI | | | | MIU | FWIU |
|---|---|---|---|---|---|---|---|
| | | Other | Chaotic | Faults | Salt | | |
| **Amplitude** | 0.5146 | 0.4604 | 0.1674 | 0.1814 | 0.3830 | 0.2981 | 0.4153 |
| **Gaussian pyramid** | 0.5232 | 0.4528 | 0.1869 | 0.2218 | 0.4064 | 0.3170 | 0.4148 |
| **Discrete wavelet** | 0.5346 | 0.4608 | 0.2353 | 0.1786 | 0.4063 | 0.3203 | 0.4229 |
| **Gabor features** | 0.7413 | 0.7070 | 0.4645 | 0.2257 | 0.4724 | 0.4674 | 0.6426 |
| **Curvelet transform** | 0.7955 | 0.7672 | 0.5128 | 0.2656 | 0.5261 | 0.5179 | 0.6997 |

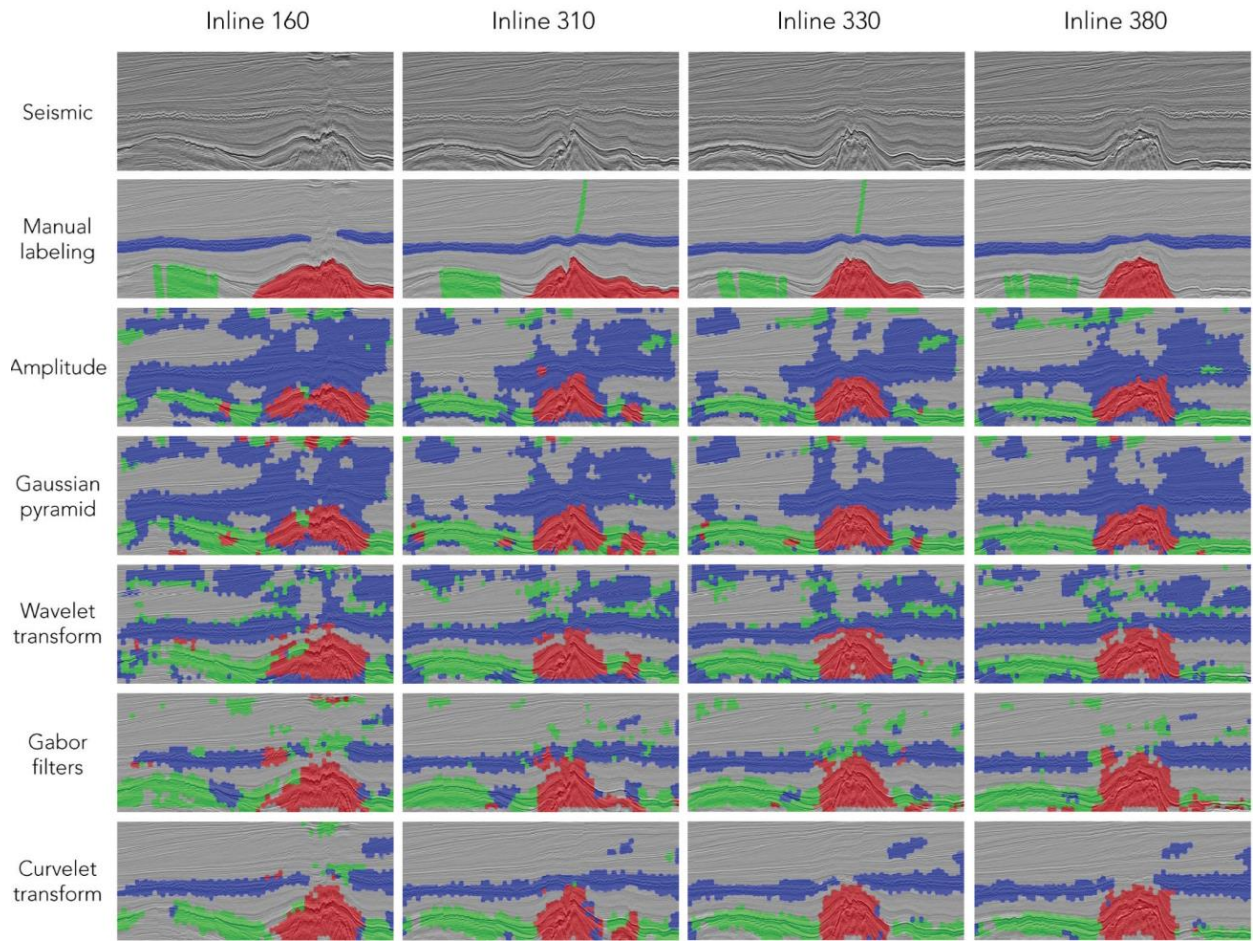

**Figure 8.** Inlines 160, 310, 330 and 380 from the Netherlands F3 block and their corresponding structure labeling using different multiresolution texture attributes in addition to manual labeling for reference. The chaotic class is blue, fault is green, salt dome is red, and other is gray. (This figure is best viewed in color)

The performance of the attributes is measured using objective measures that are commonly used in the semantic segmentation literature (Long et al., 2014). These measures compare the labels obtained using the texture attributes with the labels obtained manually shown in Figure 8. These measures are: Pixel Accuracy (PA), Intersection over Union (IU) for each class, Mean Intersection over Union (MIU) over all classes, and Frequency-Weighted Intersection over Union (FWIU). More details about these measures can be found in Appendix A.

The results of the case study show that multiresolution attributes were able to enhance the labeling workflow compared to using seismic amplitude directly. The results also suggest that directional features play a more important role than scale features for seismic image characterization. Thus, the curvelet transform and Gabor filters performed the best as they are the only two techniques with directional features. Moreover, the curvelet transform is superior to all other attributes on all metrics; mainly because of its effectiveness in representing curve-like features which constitute a large portion of the seismic section. The superiority of the curvelet attribute can be seen clearly in Figure 8.

PA and MIU assume that the four classes are represented equally in a seismic section, which is not always true. Table 2 shows the percentage of pixels that belong to every class in the four seismic sections we used in this case study. We see a clear dominance of the class labeled `other`, which contains most of the structures that are not of interest to us such as the horizons. In order to account for the varying sizes of classes, we used FWUI to compute a weighted average over the classes such that larger classes are assigned larger weights.

**Table 2.** Percentage of pixels for each class in the four inlines used in labeling.

|        | Percentage of pixels |         |        |       |
|--------|-------|---------|--------|-------|
| **Inline** | Other | Chaotic | Faults | Salt  |
| **160**  | 76.41 | 7.99    | 5.58   | 10.02 |
| **310**  | 77.23 | 8.05    | 5.75   | 8.97  |
| **330**  | 79.12 | 8.12    | 5.45   | 7.31  |
| **3830** | 79.74 | 9.38    | 4.49   | 6.39  |

The IU metric is computed for each class to provide more insight about the weaknesses or strengths of the workflow. For example, the salt dome structure was captured well in almost all multiresolution attributes because of its distinctive characteristics such as the strong edges close to the salt dome boundary. However, it is worth noting that the classifier might have learned features that are not actually associated with a given structure. For example, a large number of fault images in the training dataset have a strong reflector (see Figure 6) which is a much more dominant feature than the faults themselves, making the classifier confuse images with strong reflectors as faults. This is not a shortcoming of the attributes but rather of the nature of the labels that were used in the training. Every image is labeled with one label (class) which causes the classifier to assume that all features present in the images belong to the same class. This is not true as we have seen in the case of faults and strong reflectors. This drives the need to develop a procedure to convert image-level labels into pixel-level labels. (Alaudah and AlRegib, 2017) proposed such a procedure recently and showed promising results.

Furthermore, the varying size of the structures makes it even more difficult for the classifier to learn. For instance, faults come in different sizes as we can see in inlines number 310 and 330. These sections consist of a number of small faults and fractures at the bottom left in addition to a large fault on top of the salt dome. It appears to be difficult for the classifier to find a common and dominant feature for these two structures that belong to the same class. Thus, ideally one needs to emphasize these features using pixel-level labels.

## CONCLUSION

We examined common multiresolution texture attributes for seismic image interpretation. The attributes were examined in a seismic labeling workflow in which a seismic section is segmented and classified according to the present subsurface structure using the various multiresolution texture attributes. The results of the structure labeling showed an advantage for multiresolution attributes in improving the labeling accuracy. The curvelet transform in particular performed much better than its counterparts mainly because of its effectiveness in capturing curve-like structures which constitute large portions of seismic sections.

# ACKNOWLEDGMENTS

This work is supported by the Center for Energy and Geo Processing (CeGP) at Georgia Tech and King Fahd University of Petroleum and Minerals (KFUPM).

# APPENDIX A

# PERFORMANCE EVALUATION METRICS

If we denote $\mathcal{G}_i$ as the set of pixels manually labeled as $i$, i.e. belonging to the $i^{th}$ class, $\mathcal{F}_i$ as the set of pixels classified by our classifier as $i$, and $n_c$ as the number of classes, then the set of correctly classified pixels is the intersection set $\mathcal{F}_i \cap \mathcal{G}_i$. If we use $|*|$ to denote the number of elements in a set, then we can define the following metrics:

- Pixel Accuracy (PA) is the percentage of pixels over all classes that are correctly classified,

$$\mathrm{PA} = \frac{\sum_i |\mathcal{F}_i \cap \mathcal{G}_i|}{\sum_i |\mathcal{G}_i|}.$$

- Intersection over Union ($\mathrm{IU}_i$) is defined as the number of elements of the intersection of $\mathcal{G}_i$ and $\mathcal{F}_i$ over the number of elements of their union set,

$$\mathrm{IU}_i = \frac{|\mathcal{F}_i \cap \mathcal{G}_i|}{|\mathcal{F}_i \cup \mathcal{G}_i|}.$$

This metric measures the overlap between the two sets and it should be 1 if and only if all pixels were correctly classified. Further, we will average IU over all classes using two schemes. The first scheme is a simple average of IU over all classes defined as Mean Intersection over Union (MIU),

$$\mathrm{MIU} = \frac{1}{n_c} \sum_i \mathrm{IU}_i = \frac{1}{n_c} \sum_i \frac{|\mathcal{F}_i \cap \mathcal{G}_i|}{|\mathcal{F}_i \cup \mathcal{G}_i|}.$$

The second scheme is a weighted average of IU over all classes defined as Frequency-Weighted Intersection over Union (FWIU), in which classes with larger population are given more weight,

$$\text{FWIU} = \frac{1}{\sum_i |\mathcal{G}_i|} \sum_i |\mathcal{G}_i| \cdot \text{IU}_i = \frac{1}{\sum_i |\mathcal{G}_i|} \sum_i |\mathcal{G}_i| \frac{|\mathcal{F}_i \cap \mathcal{G}_i|}{|\mathcal{F}_i \cup \mathcal{G}_i|}.$$